\documentclass[sigconf]{acmart}

\usepackage{algorithmic}
\usepackage{textcomp}
\usepackage{url}
\usepackage{enumitem}
\usepackage{xspace}
\usepackage{soul}
\usepackage{stfloats}
\usepackage{multirow}

\copyrightyear{2026}
\acmYear{2026}
\setcopyright{cc}
\setcctype{by}
\acmConference[MSR '26]{23rd International Conference on Mining Software Repositories}{April 13--14, 2026}{Rio de Janeiro, Brazil}
\acmBooktitle{23rd International Conference on Mining Software Repositories (MSR '26), April 13--14, 2026, Rio de Janeiro, Brazil}
\acmPrice{}
\acmDOI{10.1145/3793302.3793325}
\acmISBN{979-8-4007-2474-9/2026/04}

\begin{document}

\title{Assessing Task-based Chatbots: Snapshot and Curated Datasets for Dialogflow}


\author{Elena Masserini}
\email{elena.masserini@unimib.it}
\orcid{0009-0002-6969-1500}
\affiliation{
  \institution{University of Milano-Bicocca}
  \city{Milan}
  \country{Italy}
}
\author{Diego Clerissi}
\email{diego.clerissi@unimib.it}
\orcid{0000-0001-7651-0400}
\affiliation{
  \institution{University of Milano-Bicocca}
  \city{Milan}
  \country{Italy}
}
\author{Daniela Micucci}
\email{daniela.micucci@unimib.it}
\orcid{0000-0003-1261-2234}
\affiliation{
  \institution{University of Milano-Bicocca}
  \city{Milan}
  \country{Italy}
}
\author{Leonardo Mariani}
\email{leonardo.mariani@unimib.it}
\orcid{0000-0001-9527-7042}
\affiliation{
  \institution{University of Milano-Bicocca}
  \city{Milan}
  \country{Italy}
}

\renewcommand{\shortauthors}{Masserini, et al.}

\begin{abstract}
In recent years, chatbots have gained widespread adoption thanks to their ability to assist users at any time and across diverse domains. However, the lack of large-scale curated datasets limits research on their quality and reliability. This paper presents \textsc{TOFU-D}, a snapshot of \textsc{1,788} Dialogflow chatbots from GitHub, and \textsc{COD}, a curated subset of \textsc{TOFU-D} including \textsc{185} validated chatbots. The two datasets capture a wide range of domains, languages, and implementation patterns, offering a sound basis for empirical studies on chatbot quality and security. A preliminary assessment using the Botium testing framework and the Bandit static analyzer revealed gaps in test coverage and frequent security vulnerabilities in several chatbots, highlighting the need for systematic, multi-platform research on chatbot quality and security.
\end{abstract}

\ccsdesc[500]{Software and its engineering}

\keywords{Chatbot, Dialogflow, Dataset, GitHub}


\maketitle

\section{Introduction}\label{sec:introduction}
\textit{Task-based} chatbots are ubiquitous in a multitude of domains~\cite{adamopoulou2020chatbots}, such as healthcare, bookings, and personal assistance. Different from \textit{general-purpose} chatbots (e.g., ChatGPT~\cite{chatgpt}) that do not usually target a specific domain and are not yet able to deliver mature and reliable services, 
task-based chatbots can fulfill domain-specific tasks~\cite{adamopoulou2020overview,grudin2019chatbots} and represent consolidated solutions extensively used in industry~\cite{benaddi2024systematic,adamopoulou2020chatbots}.  
Task-based chatbots can be implemented with both commercial (e.g., Dialogflow~\cite{dialogflow} and Amazon Lex~\cite{amazon-lex}) and open-source (e.g., Rasa~\cite{rasa}) platforms. All these platforms use the same set of core concepts about conversations: the \textit{intents}, which represent the goals users may want to achieve by interacting with chatbots (e.g., ordering pizza); the \textit{utterances}, which are the phrases used to train the chatbot to understand user intents; the \textit{entities}, which are the data types used in the conversations; the \textit{actions}, which are the responses and operations that the chatbot can do to address the user intents; and the \textit{flows}, which implement the possible user-bot sequence of interactions~\cite{canizares2022automating,ferdinando2024mutabot}.

The popularity and complexity of chatbots, which combine conversational data with backend logic, demand proper quality assurance techniques. 
In fact, traditional quality assurance techniques do not adapt straightforwardly to NLP-based software~\cite{cabot2021testing,li2022review,lambiase2024motivations}. 
Botium~\cite{botium} is a multi-platform test generation and execution framework for task-based chatbots. 
The framework has been exploited as a backbone in multiple studies on test case augmentation~\cite{bravo2020testing,rapisarda2025test}, scenario generation~\cite{canizares2024coverage}, and mutation testing~\cite{ferdinando2024mutabot,gomez2024mutation,clerissi2025towards}. Related studies also considered 
black-box testing~\cite{del2025automated,vasconcelos2017bottester,bozic2019chatbot}, the oracle problem~\cite{bozic2019testing,bovzic2022ontology}, runtime verification~\cite{silva2023modeling}, quality metrics~\cite{canizares2024measuring},  robustness~\cite{ruane2018botest,guichard2019assessing,iwama2019automated,liu2021dialtest}, and security concerns~\cite{bozic2020interrogating,bilika2024hello}.


A common issue with the proposed approaches is the small-scale empirical evidence that is often reported,  where subjects are selected without specific criteria, resulting in the consideration of a few and simple chatbots only. The lack of proper datasets is one of the main factors contributing to this issue. For instance, the dataset released with ASYMOB~\cite{lopez2022asymob,canizares2024measuring} consists of a collection of chatbots selected in the wild without using any specific criteria, and many of them are outdated nowadays. 
In a recent work~\cite{masserini2025brasato}, we presented \textsc{TOFU-R}, a snapshot of the Rasa chatbots present on GitHub, and \textsc{BRASATO}, a curated selection of \textsc{193} Rasa chatbots. Although useful, these datasets include chatbots developed with the same open-source platform.  

This paper addresses these limitations by presenting two new datasets with chatbots implemented for Dialogflow~\cite{dialogflow}, the popular commercial Google platform for the development of task-based chatbots~\cite{abdellatif2021comparison}. The first dataset is \textsc{TOFU-D} (\textsc{Dialogflow Task-based chatbOts From githUb}), containing a snapshot of all the Dialogflow chatbots present on Github on September 16\textsuperscript{th} 2025, (\textsc{1,788} chatbots). The second dataset is \textsc{COD} (\textsc{COllection of Dialogflow chatbots}), a curated selection of the chatbots present in \textsc{TOFU-D} (\textsc{185} chatbots). 
The two new datasets enable \emph{stronger empirical research in the area} according to multiple perspectives: 
(a) they improve \emph{generality}, since they include chatbots developed for a commercial platform that have different architecture and characteristics  (e.g., different domains, different conversational capabilities, and different complexity) than chatbots for open-source platforms, as reported in Section~\ref{sec:analysis}, (b) they increase \emph{programming language heterogeneity}, since Dialogflow chatbots are implemented with a variety of languages while Rasa chatbots are all implemented in Python, and (c) they increase the \emph{diversity} and \emph{number} of the chatbots available, by doubling the number of curated chatbots available. 

The main contributions of the paper are the \textsc{TOFU-D} and \textsc{COD} datasets for Dialogflow and the toolchain designed to create them. \textsc{TOFU-D} represents the state-of-the-practice of Dialogflow chatbots on GitHub. 
\textsc{COD} enables quality assurance research on experiments larger than prior studies, which rarely considered more than 10 chatbots~\cite{masserini2025brasato,gomez2024mutation}. The toolchain enables replicability and allows adaptation to other research goals.  
Further, we report
a preliminary experiment using testing and security tools with a selection of chatbots, which serves as an illustrative example to demonstrate the practical utility of \textsc{COD},
confirming the 
challenges already observed with other datasets, and also revealing new specific challenges.


\section{Methodology}\label{sec:methodology}

\begin{figure*}[t]
    \centering
    \includegraphics[width=0.9\textwidth]{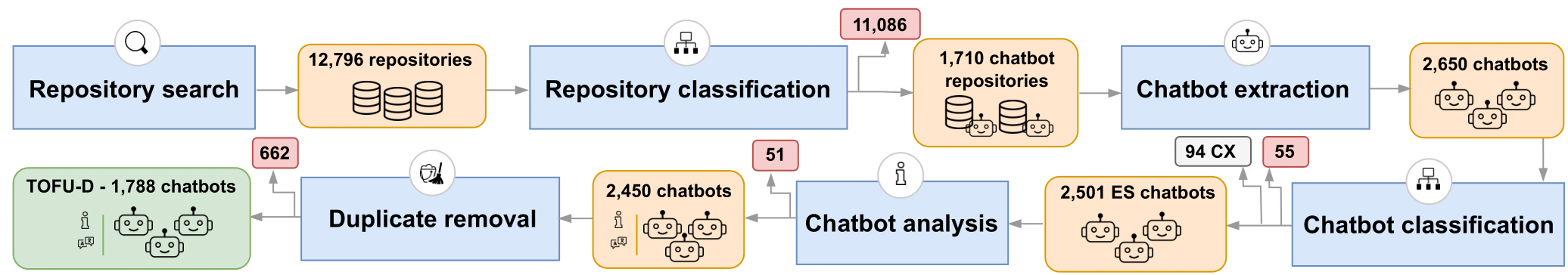}
    \caption{Methodology for the creation of the \textsc{TOFU-D} dataset.}
    \Description{The figure shows the methodology used to build the TOFU-D dataset. First, GitHub repositories potentially containing Dialogflow chatbots are collected, discarding those without a required agent file. The remaining chatbots, limited to the Dialogflow Essential free platform, are extracted and analyzed to gather conversational and functional metadata, and duplicates are removed. The final dataset consists of 1,788 chatbots.}
    \label{fig:tofu-d-methodology}
\end{figure*}

Here we discuss the methodology adopted to create the \textsc{TOFU-D} and the \textsc{COD} datasets for Dialogflow. 
The methodology is \textit{fully automated} and can be generalized to other task-based frameworks~\cite{canizares2022automating}. 

\subsection{Extraction of the \textsc{TOFU-D} dataset} 
Figure~\ref{fig:tofu-d-methodology} shows the methodology used to create the \textsc{TOFU-D} dataset.

\subsubsection*{Repository search:}We identified publicly available GitHub repositories that may include Dialogflow chatbots. We used the GitHub repository API~\cite{githubapi} to collect all the repositories that contain the keyword "Dialogflow" and at least one between "chatbot" and "agent" in the name, README file, description, or list of topics. This search resulted in 12,883 repositories. To obtain a permanent reference to the chatbot version available on Sept. 16\textsuperscript{th} 2025, we stored the latest commit SHA of each repository. This process revealed 87 empty repositories that we discarded, resulting in 12,796 non-empty ones.

\subsubsection*{Repository classification:} The occurrence of the searched keywords in a repository does not ensure the presence of an actual chatbot, since other repositories may also include those terms (e.g., repositories with chatbot development frameworks or testing tools). 
To confirm the presence of chatbots in the repositories, 
we downloaded the repositories and checked for the presence of the \textit{agent file}, which is a configuration file required by Dialogflow chatbots. 
This step excluded 11,086 repositories without chatbots, resulting in 2,650 agent files in 1,710 repositories.

\subsubsection*{Chatbot extraction:} Similarly to previous studies~\cite{masserini2025brasato}, we assume that distinct agent files in distinct folders indicate distinct agents. 
Since all agent files within the same repository were in distinct folders, we confirmed the availability of the 2,650 Dialogflow chatbots.

\subsubsection*{Chatbot classification:} Google Dialogflow chatbots can be designed to run either on Dialogflow Essentials (ES), a free platform intended for individual developers, or Dialogflow Customer Experience (CX), the platform for enterprise-scale systems. In this work, we targeted Dialogflow chatbots that can be executed without requiring a subscription. The analysis of the agent files revealed the presence of 94 CX chatbots and 2,501 ES chatbots, while 55 chatbots were discarded due to an incorrect agent file structure.

\subsubsection*{Chatbot analysis:} To provide a comprehensive description of the chatbots, we extracted the key elements defining their structure and behavior. Structural information includes the list of handled \textit{intents}, the recognized \textit{entities}, a brief \textit{description} (when available in the agent file), and the \textit{Dialogflow version}. Behavioral information includes the set of \textit{supported conversational languages}, the presence of an integration with Google Assistant~\cite{googleAssistant}, and any backend code implementing \textit{webhook services}, which are the services that are executed when a more complex operation than a plain response is needed. We also collected information on the \emph{webhook intents}, which are the intents that actually trigger the execution of webhook services, and the webhook services that are implemented through cloud functions. 
In this step, we discarded 51 chatbots due to an incorrect structure of the implementation.

\subsubsection*{Duplicate removal:} To avoid duplicates in the final \textsc{TOFU-D} dataset (e.g., due to the same chatbot code uploaded to different repositories), we checked for the presence of \textit{nearly identical chatbots}, which are chatbots that share the same list of intents, entities, webhook intents, and supported languages, and have a \textit{very similar} webhook service implementation. Specifically, we treated two chatbots as distinct only if their backend code similarity, assessed by the \texttt{difflib}~\cite{difflib} library for sequence comparison, is below 95\%, thus accounting for minor differences, such as code comments. For each group of duplicates, we selected the most representative chatbot, prioritizing the most recent Dialogflow version, to favor up-to-date subjects, higher popularity (in terms of stars and forks), and, in case of ties, the older creation date representing the original chatbot. This process led to the removal of 662 copies, resulting in the final \textsc{TOFU-D} dataset of \textsc{1,788} unique Dialogflow chatbots.

\subsection{Chatbot selection for the \textsc{COD} dataset}

\begin{figure}[H]
    \centering
\includegraphics[width=\columnwidth]{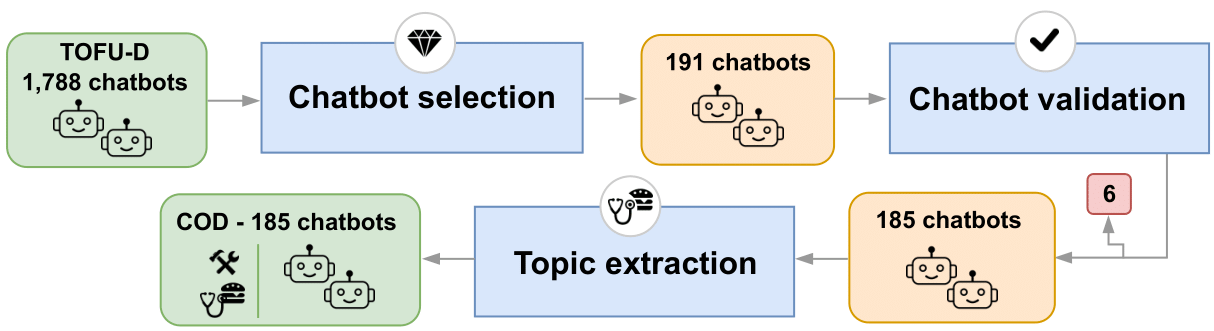}
    \caption{Methodology for the creation of the \textsc{COD} dataset.}
    \Description{The figure shows the methodology used to build the COD dataset. Chatbots are selected from TOFU-D based on minimal dialog and functional complexity and utility, validated for deployment, and labeled with topics extracted from metadata to identify their domain. The final dataset consists of 185 chatbots.}
    \label{fig:drasato-methodology}
    \vspace{-2mm}
\end{figure}

While \textsc{TOFU-D} provides a comprehensive overview of the Dialogflow chatbot population on GitHub, deriving a curated dataset with the most \textit{relevant chatbots} is important to facilitate future studies. We define as relevant the chatbots selected from 
\textsc{TOFU-D} based on three criteria that capture challenging and relevant aspects for quality assurance research~\cite{masserini2025brasato}: a minimal \textit{dialog complexity}, which requires that the chatbot can handle at least one intent and one entity; \textit{functional complexity}, which requires the chatbot to use a webhook service, available in the repository; and \textit{utility}, which requires the chatbot to support English (to be widely usable), to support the latest Dialogflow version (v2), and to be stored in a starred repository. As summarized in Figure~\ref{fig:drasato-methodology}, only 191 chatbots out of \textsc{1,788} satisfy these criteria. Each chatbot was then validated through the Dialogflow REST API~\cite{dialogflowAPI} to ensure successful deployment, which led to the removal of six chatbots due to entity configuration issues. Lastly, to identify the 
topics that are typically addressed with chatbots, 
we designed a prompt that, for each chatbot,
takes its repository title and description, its list of intents and entities, and its README files
, asking GPT-4o to associate each chatbot with a Google Play Category~\cite{googlePlayCat} to minimize variability.
To finally validate the correctness of the pipeline, an author performed a sample-based inspection of each intermediate result (e.g., to validate the topic accuracy provided by GPT-4o), not detecting any misclassification.

\section{Datasets Analysis}\label{sec:analysis}

\subsubsection*{Dialog complexity} The boxplots in Figure~\ref{fig:dialogComplexity} (a) show the dialog complexity of \textsc{TOFU-D} and \textsc{COD}, measured as the number of implemented intents and entities (using logarithmic scale). Both boxplots highlight a spread distribution in \textsc{TOFU-D}, with the number of intents and entities ranging from 1 to 316 and from 0 to 140, respectively. In particular, \textsc{TOFU-D} includes several chatbots (15\%) that do not process any information. This subset consists of both toy examples and chatbots that uniquely implement long enumerations of predefined answers to predefined user queries.  
\textsc{COD} discards these cases, as well as chatbots without any backend implementation, resulting in a more curated selection of chatbots providing actual services through conversations. Note that trivial chatbots responding according to long pre-defined enumerations may include many intents and entities; this is why \textsc{COD} does not include some apparently large, but irrelevant, chatbots.

\subsubsection*{Functional complexity} 
\begin{figure}[H] 
\vspace{-3mm}
\begin{minipage}[b]{0.53\columnwidth}
    \centering
    \includegraphics[width=\linewidth]{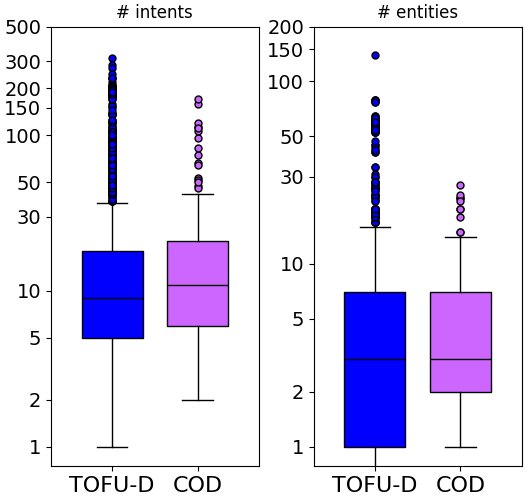}
    Dialog\\
    (a) 
    \label{fig:dialog-parameters}
\end{minipage}
\hfill
\begin{minipage}[b]{0.45\columnwidth}
    \centering
    \includegraphics[width=0.8\linewidth]{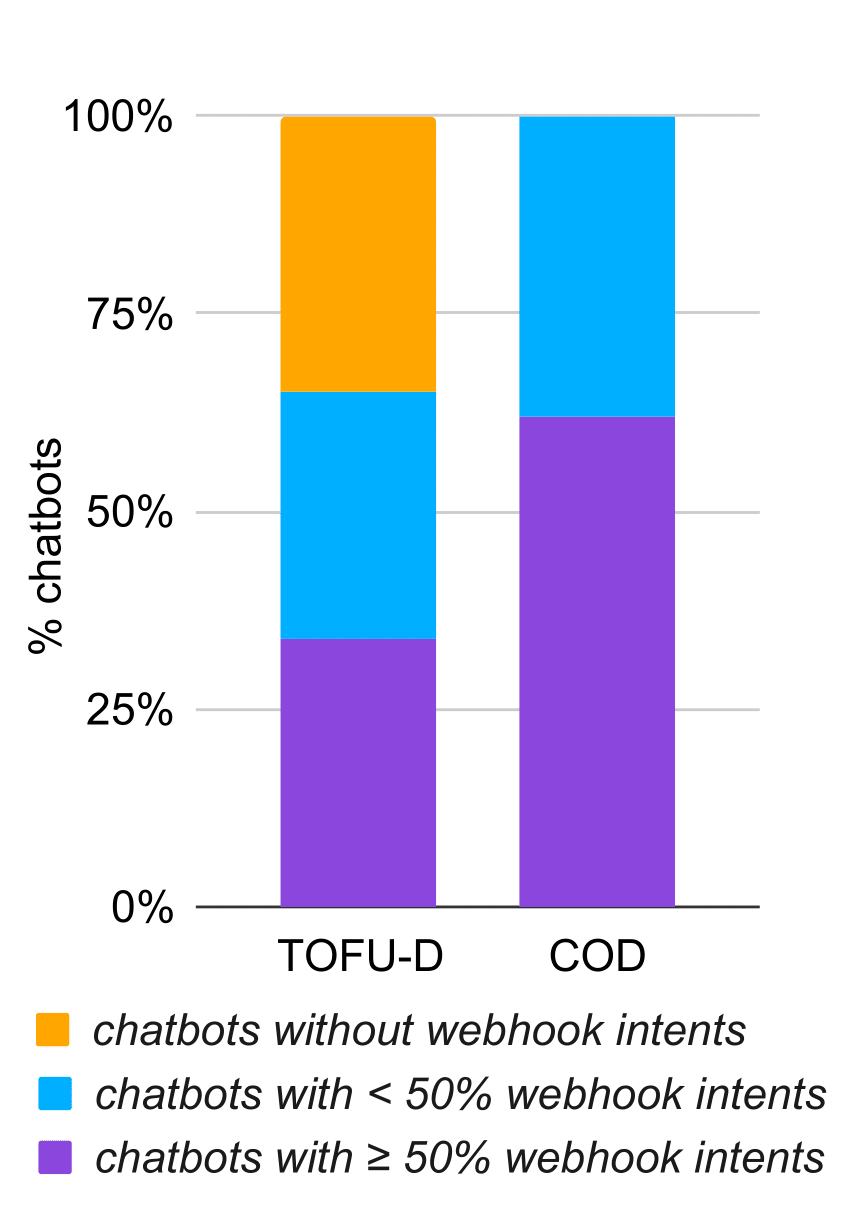}
    Functional\\
    (b)
    \label{fig:webhook-intents}
\end{minipage}
\caption{Complexity of \textsc{TOFU-D} and \textsc{COD}.} \label{fig:dialogComplexity}
\Description{The figure compares dialog and functional complexity in TOFU-D and COD. TOFU-D shows a wide distribution in the number of intents and entities, including many trivial or non-interactive chatbots, while COD excludes such cases, resulting in a more curated set of service-oriented chatbots. COD also exhibits higher functional complexity, as all chatbots include webhook-based actions.}
\vspace{-3mm}
\end{figure}

This higher complexity of \textsc{COD}, on the functional side, is highlighted by the analysis of intent types in the two datasets, as shown in Figure~\ref{fig:dialogComplexity} (b). All chatbots in \textsc{COD} implement at least one action different from the mere output of a predefined textual response  (i.e., have at least one webhook intent). Moreover, the majority of the chatbots (62\%) exploit the webhook service for more than half of the possible user intents. 

An interesting aspect of Dialogflow chatbots is their possible integration with Google's proprietary services, such as the Google Assistant and Google cloud functions. These two capabilities are relatively present in \textsc{TOFU-D} population, with only 32\% and 16\% of chatbots supporting them, respectively. Their frequency is definitely higher in \textsc{COD}, where 51\% of the chatbots support Google Assistant and 29\% of them implement cloud functions. This result confirms \textsc{COD} is representative of modern chatbots that integrate with cloud features and state-of-the-art assistants. 

\begin{figure}[h]
    \centering
    \includegraphics[width=\columnwidth]{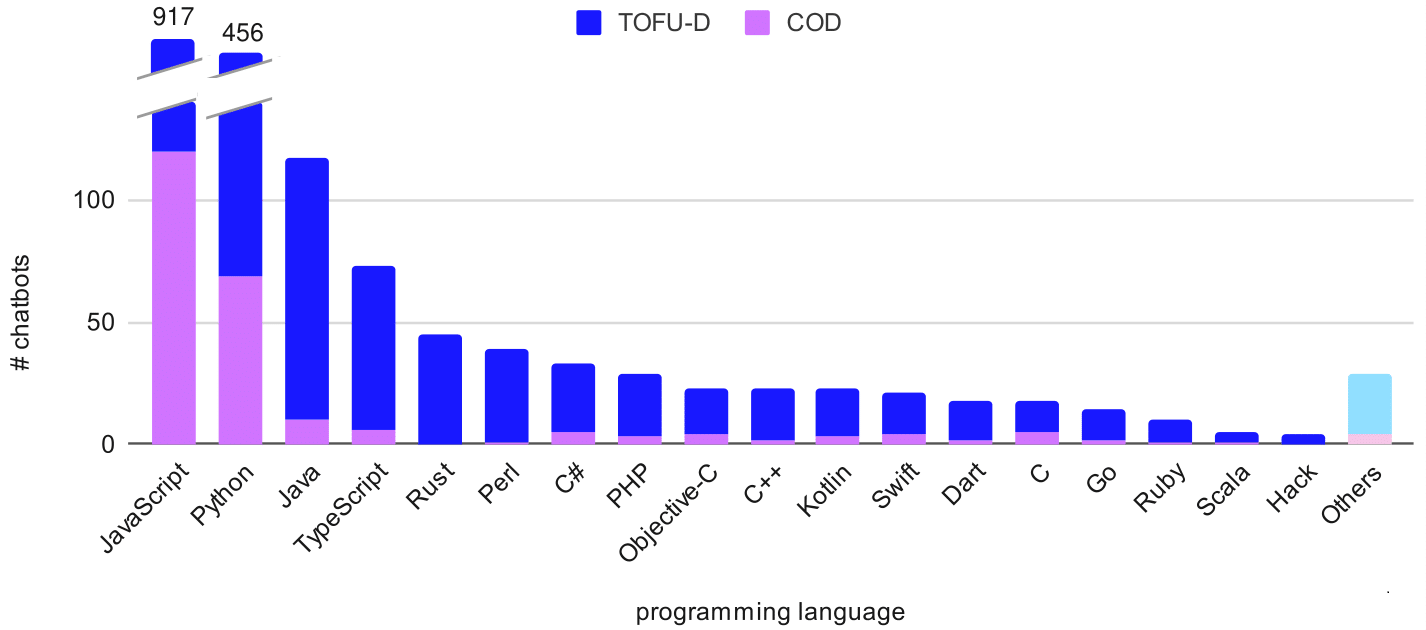}
    \vspace{-8mm}
    \caption{Programming languages in \textsc{TOFU-D} and \textsc{COD}.}
    \Description{The figure shows the number of chatbots with backend code in the most common programming languages for TOFU-D and COD datasets. Developers mainly adopt event-driven languages such as JavaScript, Python, Java, and TypeScript, with JavaScript being the most popular due to native support in Dialogflow.}
    \label{fig:prog-languages}
    \vspace{-2mm}
\end{figure}

Figure \ref{fig:prog-languages} plots the number of chatbots that include backend files in the most popular programming languages. Since chatbots are systems that must dynamically and flexibly react to user requests, developers favor programming languages with strong support to an event-driven design, like JavaScript, Python, Java, and TypeScript. JavaScript is particularly popular since it is natively supported by Dialogflow and the Google development platform. 

\subsubsection*{Utility} 
\begin{figure}[H]
    \centering
    \vspace{-2mm}
    \includegraphics[width=\columnwidth]{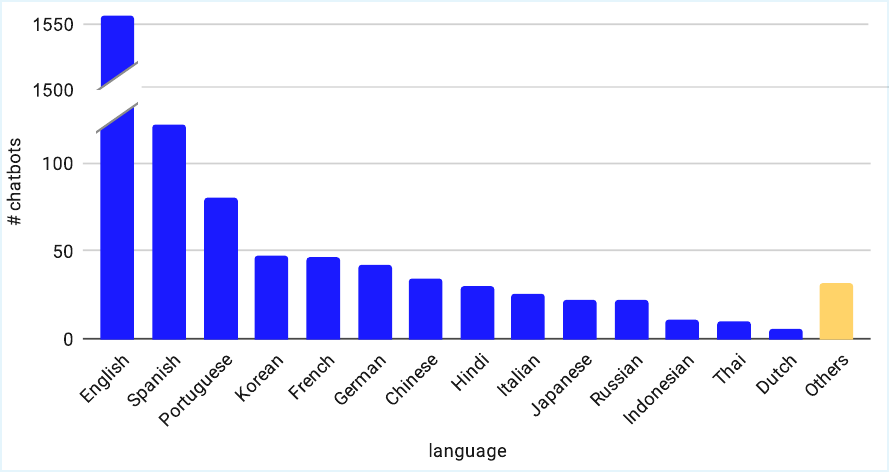}
    \vspace{-8mm}
    \caption{Conversational languages in \textsc{TOFU-D}.}
    \Description{The figure shows the languages supported by chatbots in the TOFU-D dataset, highlighting the dominance of English.}
    \label{fig:languages}
    \vspace{-2mm}
\end{figure}

Lastly, we discuss utility in terms of supported languages, Dialogflow version, and popularity of the repositories, with a final insight into the chatbot domains in \textsc{COD}. As shown in Figure~\ref{fig:languages}, the language distribution in \textsc{TOFU-D} reveals a strong predominance of English, supported by 1,557 chatbots. The set of the most popular languages is mainly consistent with the set of the most spoken languages in 2025~\cite{statista}, though with some differences in content (e.g., Italian is not one of the most spoken languages) and ranking (e.g., Chinese being less represented in \textsc{TOFU-D}). Although the standard is a monolingual implementation, a total of 159 chatbots support multiple languages, with up to 12 languages supported in three chatbots. Another common standard is the use of the latest platform version: 82\% of the chatbots in \textsc{TOFU-D} rely on Dialogflow v2, due to its availability since 2019 and the early deprecation of the previous one~\cite{dialogflowReleaseNotes}. Regarding popularity, only 37\% of the chatbots in \textsc{TOFU-D} come from starred projects on GitHub; while the number of stars for the repositories that originate the chatbots in \textsc{COD} ranges from 1 to 289. 
Chatbots in \textsc{COD} cover 28 distinct Google Play domains, with Education (18\%), Business (13\%), Medical (11\%), and Food and Drinks (7\%) being the most common topics.

\section{Early Validation with \textsc{COD}}\label{sec:uniqueness}


The datasets described in this work can support and inspire future research on task-based chatbot development and quality. \textsc{TOFU-D} and \textsc{COD} can be used to explore how correct, efficient, robust, and secure task-based chatbots are, with particular emphasis on software developed by the open-source community. Moreover, our datasets, \textsc{COD} in particular, can be used to explore the effectiveness of state-of-the-art quality assurance methods, to finally identify the research gaps that must be addressed in the future.

%




Below, we describe a preliminary investigation we conducted using \textsc{COD} about (a) the effectiveness of \emph{automatic testing} techniques, and (b) the \emph{vulnerabilities} that may affect task-based chatbots. 


To explore the effectiveness of test case generation tools, we randomly selected 10 chatbots from \textsc{COD}\footnote{The selected chatbots are: \textit{BestCovid19\_bot-DialogFlow}, \textit{ASD-Bot}, \textit{PyTacos}, \textit{Smart-School-Chatbot}, \textit{Dr.Bot}, \textit{Driving-Pal}, \textit{ChatterBot-TextClassifier}, \textit{dbpedia-chatbot-backend}, \textit{echo}, and \textit{conversational-ai-history-agent}.} and generated test cases with the state-of-the-art tool Botium~\cite{botium}, a widely used framework for the functional assessment of task-based chatbots.
We observed \textit{missing tests for intents} managing fallback behaviors (100\% of chatbots affected), greeting the user (30\%), or requiring preconditions from previous interactions to activate (30\%). We also observed cases of completely \textit{uncovered entities} (30\% of chatbots affected), or only partially covered ones (70\%). 
Furthermore, we observed cases of \textit{broken tests}, where the chatbot response was missing (10\%). Considering the results reported for Rasa chatbots~\cite{masserini2025brasato}, these preliminary findings confirm that designing effective approaches for automatic test generation is a general cross-platform challenge.


To explore how vulnerable chatbots are, we used Bandit, a static analysis tool~\cite{bandit} designed to detect security vulnerabilities in Python code. We analyzed all the 69 chatbots in \textsc{COD} that include Python code to perform a comparison with all the \textsc{193} chatbots in \textsc{BRASATO}, which are always implemented in Python.

Table~\ref{tab:security-issues} shows the six most common security issues detected in both datasets and the corresponding share of affected chatbots. 
The higher number of chatbots affected by potential vulnerabilities in \textsc{COD} is likely due to their higher complexity compared to the chatbots in \textsc{BRASATO}, which usually have a smaller backend.

Interestingly, the results show commonalities and differences between the two classes of chatbots. On one hand, missing timeouts in external APIs requests, the use of weak pseudo-random generators, improper handling of exceptions (e.g., try-catch-pass) and possible SQL injections caused by unsanitized input parameters are common to chatbots implemented with both platforms, suggesting that chatbot developers may easily overlook the integration of external APIs, databases, and random generators when implementing chatbots. 
On the other hand, Dialogflow chatbots exhibit vulnerabilities specific to their webhook-based architectures, including API misconfigurations that accept connections from any network interface (41\% of the chatbots) and that may allow the execution of arbitrary code (32\% of the chatbots). 

These results show how task-based chatbots must be studied according to a multi-platform perspective, which allows researchers to address both the general challenges that characterize the chatbot technology and the ones that are specific to certain platforms.


\begin{table}[h!]
\setlength{\tabcolsep}{2pt}
\centering
\caption{Most common security issues in \textsc{BRASATO} and \textsc{COD}.}
\Description{Table 1 shows the six most common security issues detected in BRASATO and COD datasets. Both platforms share vulnerabilities such as missing API timeouts, poor exception handling, and unsanitized inputs. }
\label{tab:security-issues}
\resizebox{0.95\columnwidth}{!}{
\begin{tabular}{lc|lc}
\toprule
\multicolumn{2}{c|}{\textbf{\textsc{BRASATO}}} & \multicolumn{2}{c}{\textbf{\textsc{COD}}} \\ 
\cmidrule(lr){1-2} \cmidrule(lr){3-4}
\textbf{Issue} & \textbf{Freq.} & \textbf{Issue} & \textbf{Freq.} \\ 
\midrule
request without timeout & 29\% & all interfaces binding & 41\% \\
pseudo-random generator & 17\%  & flask debug true  &  32\% \\
pickle deserialization & 5\%  & request without timeout & 32\% \\
hardcoded SQL string & 5\% & pesudo-random generator  &  26\% \\
silent exception handling & 5\% & silent exception handling &  15\% \\
assert ignored by compiler & 2\% & hardcoded SQL string &  10\% \\
\bottomrule
\end{tabular}
}
\vspace{-1mm}
\end{table} 

\section{Conclusions}\label{sec:conclusions}

This paper presents \textsc{TOFU-D}, a snapshot of the Dialogflow chatbots present on GitHub, and \textsc{COD}, a curated dataset of \textsc{185} Dialogflow chatbots. 
These datasets represent a step towards a more systematic quality assessment of chatbots, as they provide a benchmark of real-world subjects for the validation and development of novel strategies. In particular, \textsc{COD}: 
i) provides a collection of chatbots implemented for a different (commercial) platform compared to Rasa~\cite{masserini2025brasato}, mitigating the risk of developing research that overfits a single technology, ii) provides chatbots implemented in multiple programming languages, iii) doubles the total number of curated chatbots available (\textsc{185} in \textsc{COD}, 193 in \textsc{BRASATO}) for experimentation, iv) includes chatbots with unique features, such as cloud functions and Google Assistant technology, v) includes chatbots with testing and security challenges complementary to existing datasets.
Future work concerns the use of the datasets to systematically develop and experiment with quality assurance methods. 



\noindent \emph{Data Availability}. The datasets and the scripts are publicly available at \url{https://gitlab.com/securitychatbot/dialogflow-chatbot-dataset}.

\bibliographystyle{ACM-Reference-Format}
\bibliography{references}

\end{document}